\documentstyle[preprint,aps,epsf,floats]{revtex}                            
\begin{document}
\tighten

\preprint{\vbox{\hbox{JHU--TIPAC--96022}
\hbox{hep-ph/9611321}}}

\title{Testing Quarkonium Production with Photoproduced $J/\psi + \gamma$}
\author{Thomas~Mehen}
\address{Department of Physics and Astronomy,
The Johns Hopkins University\\
3400 North Charles Street,
Baltimore, Maryland 21218 U.S.A.\\
{\tt mehen@dirac.pha.jhu.edu}}

\date{November 1996}

\maketitle
\begin{abstract}

I compute the leading color-octet contributions to the process $\gamma + p \rightarrow J/\psi + \gamma~(+ X)$ within the non-relativistic QCD (NRQCD) factorization formalism. In the color-singlet model, $J/\psi + \gamma$ can 
only be produced when the photon interacts through its structure function,
while the color-octet mechanism allows for production of $J/\psi + \gamma$ 
via direct photon-gluon fusion. Resolved photon processes can be easily be distinguished from direct photon processes by examining the fraction of the incident photon energy carried away by the $J/\psi$ in the event. Therefore, this process provides a conclusive test of the color-octet mechanism.  
$J/\psi + \gamma$ production is particularly sensitive to the NRQCD matrix element which figures prominently in the fragmentation production of 
$J/\psi$ at large $p_{\perp}$ in hadron colliders.  I also examine the predictions of the color evaporation model (CEM)  of quarkonium production 
and find that this process can easily discriminate between the NRQCD factorization formalism and the CEM.
 
\end{abstract}

\pagebreak

In recent years, quarkonium production has become the subject of intense theoretical and experimental investigation. This has been motivated by the observation of gross discrepancies between experimental measurements of $J/\psi$ and $\psi^{\prime}$ production at CDF \cite{CDF}, and theoretical calculations based on the Color-Singlet Model (CSM).  Attempts to understand this discrepancy focus on new production mechanisms which allow the $c \overline{c}$ to be produced in a color-octet state and evolve nonperturbatively into charmonium. In this paper, I examine the 
photoproduction of associated $J/\psi + \gamma$ within two different 
frameworks for computation of quarkonium production: the non-relativistic 
QCD (NRQCD) factorization formalism and the color evaporation model (CEM).  
In the NRQCD factorization formalism, non-perturbative aspects of quarkonium production are organized in an expansion in powers of $v$, the relative velocity of the $Q\overline{Q}$ in quarkonia. For production of S-wave quarkonia, the CSM emerges in the non-relativstic limit $v \rightarrow 0$, 
and color-octet mechanisms are the leading order corrections in the $v$ expansion. In the CEM, the hadronization of $Q \overline{Q}$ pairs into quarkonia is assumed to be dominated by long distance fluctuations of gluon fields which motivates a statistical treatment of color.

At leading order, the CSM predicts that $J/\psi + \gamma$ can be produced in photon-proton collisions only if the incident photon is resolved, resulting 
in $J/\psi$ that carry a small fraction of the incident photon energy. Color-octet mechanisms allow $J/\psi + \gamma$ to be produced through direct photon-gluon fusion and the resulting $J/\psi$ are typically much more energetic. The energy spectra of $J/\psi$ then directly indicates the relative importance of color-octet and color-singlet production mechanisms.  Since the CSM, the NRQCD factorization formalism and the CEM each differ in the relative importance of the octet and singlet components, the $J/\psi$ energy spectra predicted by the three models differ in a qualitative way. The leading order calculations in this paper will be subject to considerable theoretical uncertainties due to higher order QCD corrections. However, for reasons that will be discussed later, I expect the shape of the energy spectra to be well predicted by the leading order calculation even though the normalization of 
the total cross section is not. This makes $\gamma + p \rightarrow J/\psi + \gamma~(+~X)$ an excellent process for testing  theories of quarkonium production. 

Before I discuss theoretical predictions for $\gamma + p \rightarrow J/\psi + \gamma~(+~X)$, I will briefly review the deficiencies of the CSM and motivate the NRQCD factorization formalism and the CEM. A comprehensive review of 
recent developments in the theory of quarkonia production can be found in ref.\cite{BFY}.

In the CSM\cite{Schuler}, quarkonium is viewed as a non-relativistic, color-singlet bound state of a $Q \overline{Q}$ with definite angular 
momentum quantum numbers. Production (or decay) rates for a quarkonium state are computed by calculating the production (or decay) of a $Q \overline{Q}$ pair which is constrained to be in a color-singlet state and have the same angular momentum quantum numbers as the physical quarkonium. The dynamics of the formation of the bound state are parametrized by the wavefunction, or its derivatives, evaluated at the origin. 

Such a description is not entirely satisfactory from either a theoretical or 
an experimental point of view. For instance, a calculation of inclusive hadronic decays of P-wave quarkonia in the CSM reveals infrared divergences 
at $O(\alpha_s^3)$ \cite{Barbieri}.  These divergences cannot be simply absorbed into the single existing non-perturbative parameter, the wave 
function at the origin. Furthermore, recent calculations of quarkonium production in the CSM fall below the observed production of $\psi^{\prime}$ 
at CDF by a factor of 30 \cite{BDFM}. These developments have led to a new paradigm for the computation of quarkonia production and decay based on the non-relativistic QCD (NRQCD) factorization formalism of Bodwin, Braaten and Lepage\cite{BBL2}. In this formalism, the cross section for production of a quarkonium state $H$ is written as:
\begin{equation}\label{Factor}
\sigma(H) = \sum_n c_n \langle 0|O_n^H|0\rangle .
\end{equation} 
(A similar expression holds for quarkonium decay rates.) The short-distance coefficients $c_n$ are computable in perturbation theory.  The $\langle 0|O_n^H|0 \rangle$ are matrix elements of NRQCD operators of the form:  
\begin{equation}\label{Matrix}
\langle 0|O_n^H|0 \rangle = \sum_X \sum_{\lambda} \langle 0|\kappa_n^{\dagger}|H(\lambda) + X\rangle \langle H(\lambda) + X |\kappa_n|0 \rangle.
\end{equation}
The $\kappa_n$ in eqn.~(\ref{Matrix}) is a bilinear in heavy quark fields 
which creates a $Q \overline{Q}$ pair in a state with definite color and angular momentum quantum numbers.  These matrix elements describe the 
evolution of the $Q \overline{Q}$ into a final state containing the 
quarkonium $H$ plus additional hadrons ($X$) which are soft in the 
quarkonium rest frame. Since this paper will only deal with unpolarized production, a sum over polarizations ($\lambda$) of the quarkonium is 
included in eqn.~(\ref{Matrix}). Below I will use a shorthand notation in 
which the matrix elements are given as $\langle O^H_{(1,8)}(^{2S+1}L_J) \rangle$. The angular momentum quantum numbers of the $Q\overline{Q}$ 
produced in the short-distance process are given in standard spectroscopic notation, and the subscript refers to the color configuration of the $Q\overline{Q}$: $1$ for a color singlet and $8$ for a color octet. These matrix elements obey simple scaling laws with respect to $v$, the relative velocity of the $Q \overline{Q}$. This allows for a systematic expansion in 
the two small parameters relevant for quarkonia: $\alpha_s(2 m_Q)$ and $v$. 
For production of S-wave quarkonia, the results of the CSM are recovered in 
the limit $v \rightarrow 0$, since the leading matrix element in eqn.~(\ref{Factor}) can be related to the square of the wavefunction at the origin in the CSM. 

This formalism solves the problem of infrared divergences because the 
infrared divergences in the calculation of the short-distance coefficients 
can be systematically factored into the matrix elements appearing in eqn.~(\ref{Factor}). Thus a clean separation of long and short-distance 
scales is obtained. Furthermore, new quarkonium production mechanisms are now possible, since it is no longer required that the $Q \overline{Q}$ produced 
in the short-distance process have the same color and angular momentum quantum numbers as the quarkonium state. Central to the resolution of the $\psi^{\prime}$ anomaly \cite{BF} are color-octet processes, in which the $Q \overline{Q}$ is produced in a color-octet state in the short-distance 
process. For S-wave quarkonia production, the color-octet matrix elements 
are suppressed by $v^4$ relative to the leading color-singlet matrix element. This is not a very large suppression in the case of charmonium since one expects $v^2 \simeq 0.25$. Furthermore, in some instances, the short-distance process which produces a color-octet $Q \overline{Q}$ can be lower order in 
the strong coupling than the corresponding color-singlet process. In this 
case, color-octet and color-singlet mechanisms are equally important since 
the color-octet is suppressed by a factor of $\sim v^4 \pi/\alpha_s \sim 
O(1)$.

The factorization theorem for production of quarkonia is a prediction of QCD which has been proven at a level of rigor similar to proofs of factorization for the Drell-Yan process in hadron collisions. Its utility as a tool for quantitative description of quarkonia production depends on the convergence 
of the expansion in the parameter $v$. If the expansion in $v$ is sufficiently convergent, then only a small number of the non-perturbative matrix elements 
in eqn.~(\ref{Factor}) need to be retained, and the formalism is predictive as well as rigorous.  It is possible that for charmonium the velocity expansion 
is not well behaved, in which case eqn.~(\ref{Factor}) would cease to be of practical value. 

A phenomenological alternative to the NRQCD formalism which has recieved renewed attention in the wake of the $\psi^{\prime}$ anomaly is the color evaporation model (CEM)\cite{ColorEvap}. The CEM is a simple, intuitive model for quarkonia production which relies on a minimal number of undetermined parameters. Like the NRQCD factorization formalism, the CEM allows for color-octet production, since no constraints are placed on the color configuration of the quarks produced in the short-distance process. The CEM differs from the NRQCD formalism in its treatment of non-perturbative aspects of quarkonia production. In this model, the hadronization of the $Q\overline{Q}$ pair is charcterized by multiple soft gluon emission which is not suppressed by any power of $v$, and so the treatment of color is statistical. The sum of cross sections for production of all quarkonium states is related to the differential cross section for producing quark-anti-quark pairs by:
\begin{equation}\label{CEonium}
\sigma_{onium} =  {1 \over 9} \int_{2m_c}^{2m_D} dm {d \sigma_{c\overline{c}}. \over dm}
\end{equation}
A similar formula holds for the production of open charm:
\begin{equation}\label{CEopen}
\sigma_{open} =  {8 \over 9} \int_{2m_c}^{2m_D} dm {d \sigma_{c\overline{c}} \over dm} + \int_{2m_D} dm {d \sigma_{c\overline{c}} \over dm} .
\end{equation}
The variable $m$ is the invariant mass of the $c\overline{c}$ pair.  The 
factor $1/9$ represents the probability that the $c\overline{c}$ pair emerges in a color-singlet state. For production of exclusive quarkonium states more non-perturbative information is needed. In the CEM, it is assumed that the fraction of quarkonium states that emerge as $J/\psi$ is independent of the particular production process and can be described by a single  parameter:
\begin{equation}
\sigma_{J/\psi} = \rho_{J/\psi} \sigma_{onium} .
\end{equation}

In this paper, I will compute the production cross sections for $\gamma + p \rightarrow J/\psi + \gamma~(+~X)$ in the NRQCD factorization formalism and 
the CEM. Both calculations will be to leading order in $\alpha_s$. The 
absolute normalization of these leading order calculations is not to be 
trusted to a high degree of precision because of large theoretical uncertainties due to ambiguity in the choice of the charm quark mass, scale setting ambiguities, and potentially large higher order QCD corrections. Therefore measurement of the total cross section will not be a sensitive test of quarkonium production theory until the substantial theoretical 
uncertainties are reduced. However, differential distributions are expected 
to be less sensitive to these uncertainties. As an example, the higher order QCD corrections to color-singlet
photoproduction of $J/\psi$ \cite{K} increase the total cross section by a factor of 2 but leave the differential cross sections $d\sigma/dz$ and $d\sigma/dp_{\perp}$ essentially unaffected. Similarly, leading order computations \cite{ColorEvap} within the CEM show that the model correctly produces the $p_{\perp}$ dependence of $J/\psi$ production at the Tevatron 
but a K-factor of roughly 2 is needed to achieve correct normalization of the total cross section.

I first discuss computation of $J/\psi + \gamma$ production cross sections 
within the NRQCD factorization formalism. The short-distance coefficients appearing in eqn.~(\ref{Factor}) can be computed by matching a perturbative calculation in full QCD with a corresponding perturbative calculation in 
NRQCD. The production cross section of a $c \overline{c}$ pair with relative three momentum {\bf q} is computed in full QCD and Taylor expanded in powers 
of {\bf q}.  In this Taylor expansion, the four-component Dirac spinors are expressed in terms of non-relativistic two-component heavy quark spinors. The NRQCD matrix elements on the right hand side of eqn.~(\ref{Factor}) are easily expressed in terms of the two-component heavy quark spinors and powers of {\bf q}, and then the $c_n$ appearing in eqn.~(\ref{Factor}) are chosen so that the full QCD calculations and NRQCD calculations agree. A more detailed discussion of the matching procedure, along with examples of specific cross section calculations, is found in ref.\cite{BC}. Here I give the parton level differential cross sections relevant for resolved and direct photoproduction 
of $J/\psi + \gamma$, including both color-singlet and color-octet contributions:

\begin{equation}\label{GaGPsiGa8}
{d\sigma \over dt}(\gamma + g \rightarrow J/\psi + \gamma) = {64\pi^2 \over 3} {e_c^4 \alpha^2 \alpha_s  m_c \over  s^2}\left( {s^2 s_1^2 + t^2 t_1^2 + u^2 u_1^2 \over s_1^2 t_1^2 u_1^2} \right)\langle O_8^{J/\psi}(^3S_1)\rangle ,
\end{equation}

\begin{eqnarray}\label{GGPsiGa}
{d\sigma \over dt}(g + g \rightarrow J/\psi + \gamma) = {\pi^2 e_c^2 \alpha \alpha_s^2 m_c\over  s^2} \left[ {10\over 9} \left( {s^2 s_1^2 + t^2 t_1^2 + u^2 u_1^2 \over s_1^2 t_1^2 u_1^2} \right)
\langle O_8^{J/\psi}(^3S_1)\rangle \right. \nonumber \\
+ {16\over 27} \left( {s^2 s_1^2 + t^2 t_1^2 + u^2 u_1^2 \over s_1^2 t_1^2 u_1^2} \right) \langle O_1^{J/\psi}(^3S_1)\rangle + {3\over 2} {t u \over s s_1^2 m_c^2} \langle O_8^{J/\psi}(^3S_1)\rangle \nonumber \\
+ \left.{3\over 2}  {1 \over s s_1^2 m_c^4} \left(2 s (2 m_c)^2 + 3 t u - {4 t u (2 m_c)^2 \over s_1} \right) \langle O_8^{J/\psi}(^1P_0)\rangle  \right] ,
\end{eqnarray}

\begin{equation}\label{qqPsiGa}
{d\sigma \over dt}(q + \overline{q} \rightarrow J/\psi + \gamma) = {4 \pi^2 \over 9} {e_q^2 \alpha \alpha_s^2 \over s^2 (2 m_c)^2} \left( {t \over u} + {u \over t} + {2 (2 m_c)^2 s \over t u } \right)
{\langle O_8^{J/\psi}(^3S_1)\rangle \over m_c} .
\end{equation}
In these formulae, $s$,$t$, and $u$ are the Mandelstam variables; $s_1 = s - 4 m_c^2$, $t_1 = t - 4 m_c^2$, and  $u_1 = u - 4 m_c^2$. 

Two variables which are more commonly used in photoproduction are $z$ and $p_{\bot}$.  The variable $z$ is defined to be $p_{\psi}\cdot p_N /p_{\gamma}\cdot p_N$. The four-momenta of the initial state photon, the $J/\psi$, and the proton are denoted $p_{\gamma}$, $p_{\psi}$ and $p_N$, respectively. In a frame in which the proton is at rest, $z$ represents the fraction of the photon energy carried away by the $J/\psi$.  $p_{\bot}$ is 
the magnitude of the $J/\psi$ three-momentum normal to the beam axis. The differential cross section for unresolved photoproduction given above is expressed in terms of these variables by making the substitutions:
\begin{eqnarray}\label{sub}
s = {p_{\bot}^2 + (2 m_c)^2(1 - z) \over z (1 - z)},~ \nonumber \\  
t = -{p_{\bot}^2 + (2 m_c)^2(1 - z) \over z}, \nonumber \\
u = -{p_{\bot}^2 \over 1 - z}.~~~~~~~~~~
\end{eqnarray}
The total cross section is obtained by convolving the differential cross section with the gluon distribution function of the proton, $f_{g/p}(x)$:
\begin{equation}
\sigma_{tot} = \int dx dt f_{g/p}(x) {d\sigma \over dt} = \int dz dp_{\bot}^2 {x f_{g/p}(x) \over z(1 - z)} {d\sigma \over dt}.
\end{equation}
The momentum fraction of the gluon, $x$, is related to $z$ and $p_{\perp}$ by:
\begin{equation}
x = {s \over s_{\gamma p} }= {p_{\bot}^2 + (2 m_c)^2(1 - z) \over s_{\gamma p} z (1 - z)},
\end{equation}
where $s_{\gamma p}$ is the photon-proton center of mass energy squared.
For processes involving a resolved photon the formulae in eqn.~(\ref{sub}) are correct provided $z$ is replaced with $\overline{z} = z/y$, where $y$ is the fractional momentum carried by the parton coming from the photon. The total cross section is then obtained by integrating over the parton distribution functions of the proton and photon:
\begin{equation}
\sigma_{tot} = \int dx dy dt f_{g/p}(x) f_{g/\gamma}(y) {d\sigma \over dt} = \int dy dz dp_{\bot}^2 {x f_{g/p}(x) f_{g/\gamma}(y)\over y \overline{z}(1 - \overline{z})} {d\sigma \over dt}.
\end{equation}
In this case, $x = s/y s_{\gamma p}$. Similar formulae hold for processes in which the initial state partons are quarks.

For calculation within the CEM, the matrix elements for $\gamma + g 
\rightarrow c + \overline{c} + \gamma$, $g + g \rightarrow c + \overline{c} + \gamma$ and $q + \overline{q} \rightarrow c + \overline{c} + \gamma$ are found in refs.\cite{GK,EK}. I fix $\rho_{\psi}$ to be 0.45, which is consistent with the results of ref.\cite{ColorEvap}. A very similar computation of $e + p \rightarrow J/\psi + \gamma$ within the CEM can be found in ref.\cite{KR}.
 
For both CEM and NRQCD calculations, I use the GRV leading order parton distributions for the proton \cite{GRVproton} and the photon \cite{GRVphoton}. I use the one loop expression for the running coupling with $\Lambda_{QCD} = 232~{\rm GeV}$ and $n_f =3$. (This is the definition of $\alpha_s$ used in the GRV parton distribution functions.) The scale for the strong coupling, $\mu$, is given by $\mu^2 = p_{\perp}^2 + (2m_c)^2$, where $m_c = 1.5~{\rm GeV}$. For numerical values of the NRQCD matrix elements I use:
\begin{eqnarray}
\langle O_1^{J/\psi}(^3S_1) \rangle =  1.1~{\rm GeV}^3~~\nonumber \\
\langle O_8^{J/\psi}(^3S_1) \rangle =  0.0066~{\rm GeV}^3 \nonumber \\
\langle O_8^{J/\psi}(^1S_0) \rangle =  0.04~{\rm GeV}^3~~ \nonumber \\
\langle O_8^{J/\psi}(^1S_0) \rangle =  - 0.003~{\rm GeV}^3
\end{eqnarray}
The matrix element $\langle O_1^{J/\psi}(^3S_1) \rangle$ is extracted from the leptonic $J/\psi$ decay in ref.\cite{BSK}. In ref. \cite{CL}, the value of  $\langle O_8^{J/\psi}(^3S_1) \rangle$ is determined from a fit to charmonium production measured at CDF. The values of $\langle O_8^{J/\psi}(^1S_0) \rangle$ and $\langle O_8^{J/\psi}(^3P_0) \rangle$ are determined from a study of electroproduction of $J/\psi$ found in ref.\cite{S1}. The electroproduction results are consistent with linear combinations of $\langle O_8^{J/\psi}(^1S_0) \rangle$ and $\langle O_8^{J/\psi}(^3P_0) \rangle$ extracted at CDF\cite{CL}, low energy fixed target experiments \cite{BR} and photoproduction\cite{AFM}. The extractions of color-octet NRQCD matrix elements use leading order theoretical calculations, and hence the numerical values listed above should 
be regarded as rough estimates rather than exact results.

The negative value for $\langle O_8^{J/\psi}(^3P_0)\rangle$ may at first sight seem unphysical because the tree level definition of this operator is positive definite. This could be taken as a sign that the leading order extractions of the NRQCD matrix elements are not be trusted. However, in ref.\cite{S1} it is argued that loop corrections may actually make the matrix elements of the renormalized operator negative. The operator $O_8^{J/\psi}(^3P_0)$ recieves a quadratically divergent loop correction proportional to the leading color-singlet operator $O_1^{J/\psi}(^3S_1)$. Perturbative consistency may suggest neglecting such an effect in a leading order calculation because this mixing is formally higher order in $\alpha_s$. However, the mixing is enhanced by a factor $1/v^4$ since the operator $O_1^{J/\psi}(^3S_1)$ is lower order in the velocity expansion. In any case, photoproduction of $J/\psi + \gamma$ is not very sensitive to $\langle O_8^{J/\psi}(^3P_0)\rangle$, since this matrix element only appears in resolved color-octet processes whose contibution is heavily suppressed. 

\begin{figure}
\epsfxsize=10cm
\hfil\epsfbox{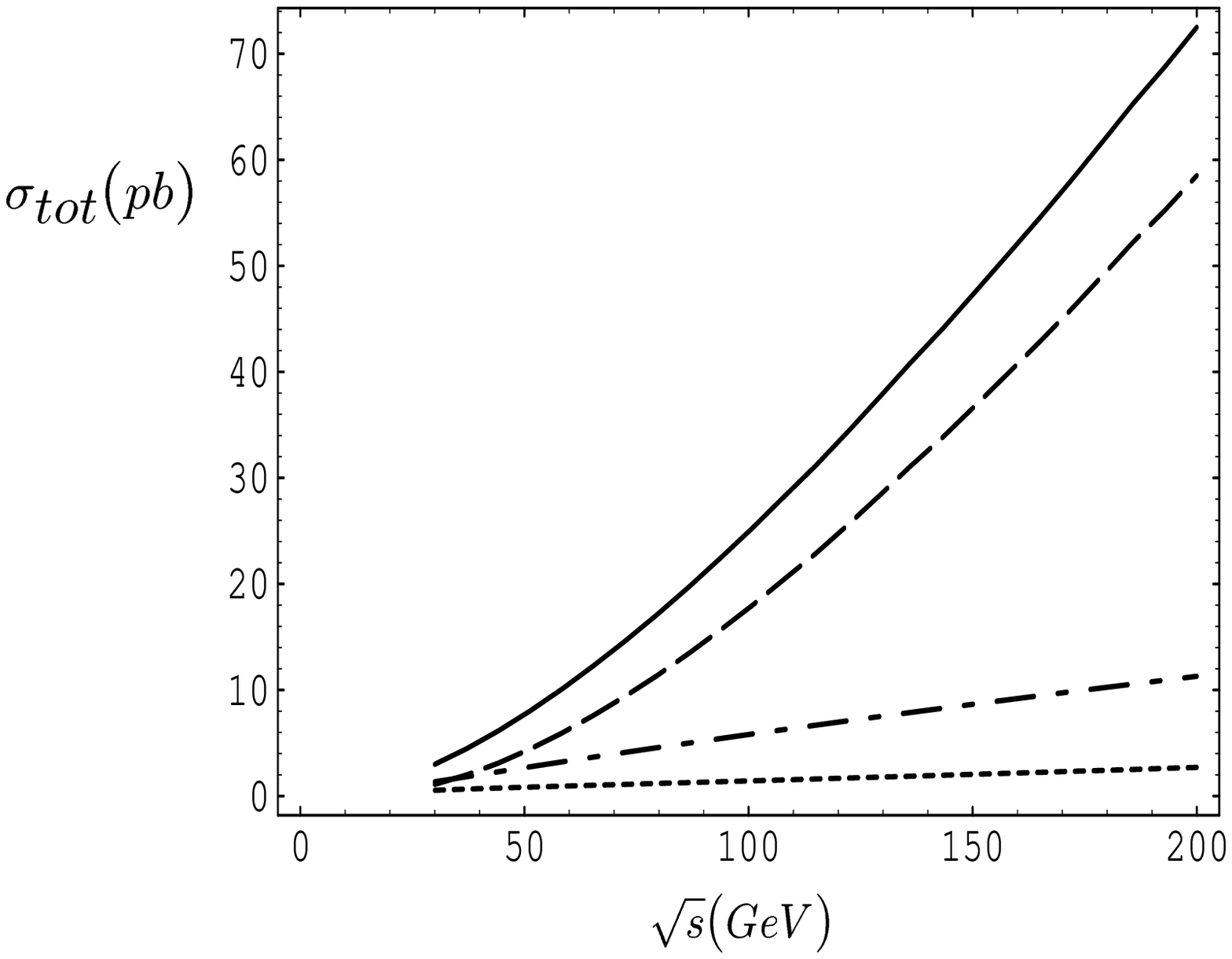}\hfill
\caption{} 
\label{NRQCDTot}
\end{figure}

\begin{figure}
\epsfxsize=10cm
\hfil\epsfbox{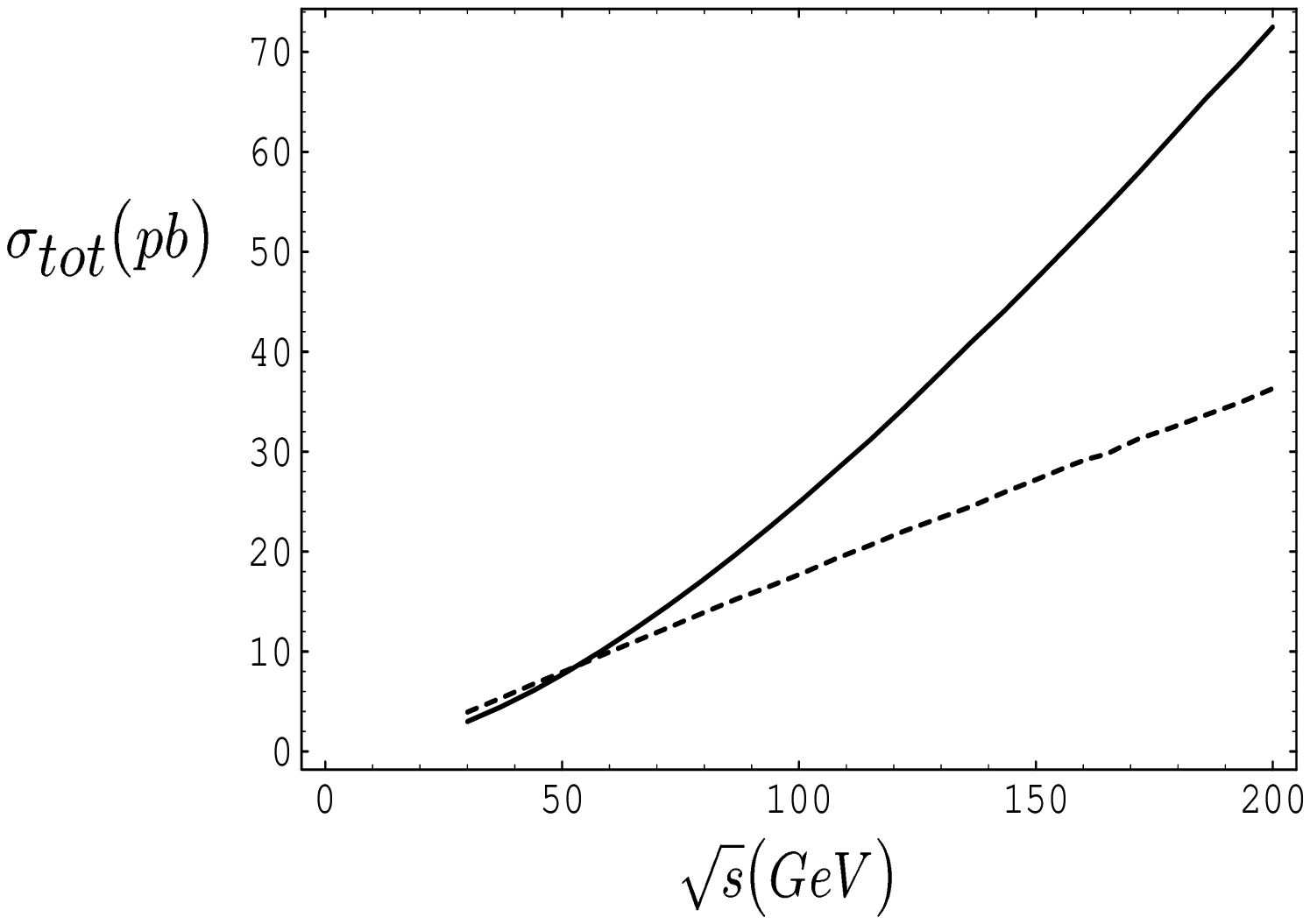}\hfill
\caption{} 
\label{Tot}
\end{figure}

I compute the total cross section for $\gamma + p \rightarrow J/\psi + \gamma~(+ X)$ for the range of energies typical of the HERA collider, $30~{\rm GeV} < \sqrt{s} < 200~{\rm GeV}$. The cuts  $z < 0.9$ and $p_{\perp} > 1~{\rm GeV}$ are imposed in order to avoid contamination due to higher twist effects and diffractive processes. The theoretical prediction of the NRQCD factorization formalism is shown in fig.~\ref{NRQCDTot}. Resolved color-singlet processes give the leading contribution to the total cross section. Direct color-octet processes give a contribution that is between 20-50\% of the total cross section for most values of $\sqrt{s}$. Resolved color-octet contributions are negligible.  Associated $J/\psi + \gamma$ production is only sensitive to the color-octet matrix element $\langle O_8^{J/\psi}(^3S_1) \rangle$ and very insensitive to $\langle O_8^{J/\psi}(^1S_0) \rangle$ and $\langle O_8^{J/\psi}(^3P_0) \rangle$. In fig.~\ref{Tot}, I compare the NRQCD and CEM predictions for the total cross section. At most values of $\sqrt{s}$ the CEM prediction is lower than the NRQCD prediction, but never by more than a factor of 2. 

I also compute $d\sigma/dp_{\perp}^2$ for the NRQCD factorization formalism and the CEM. The shapes of the $p_{\perp}$ distribution of the resolved color-singlet, direct color-octet and resolved color-octet components of the NRQCD predictions are all very similar. The CEM $p_{\perp}$ distribution differs from NRQCD only in the overall magnitude. Since the differences in magnitude are already reflected in the total cross section, I do not show the $p_{\perp}$ distributions.

\begin{figure}
\epsfxsize=10cm
\hfil\epsfbox{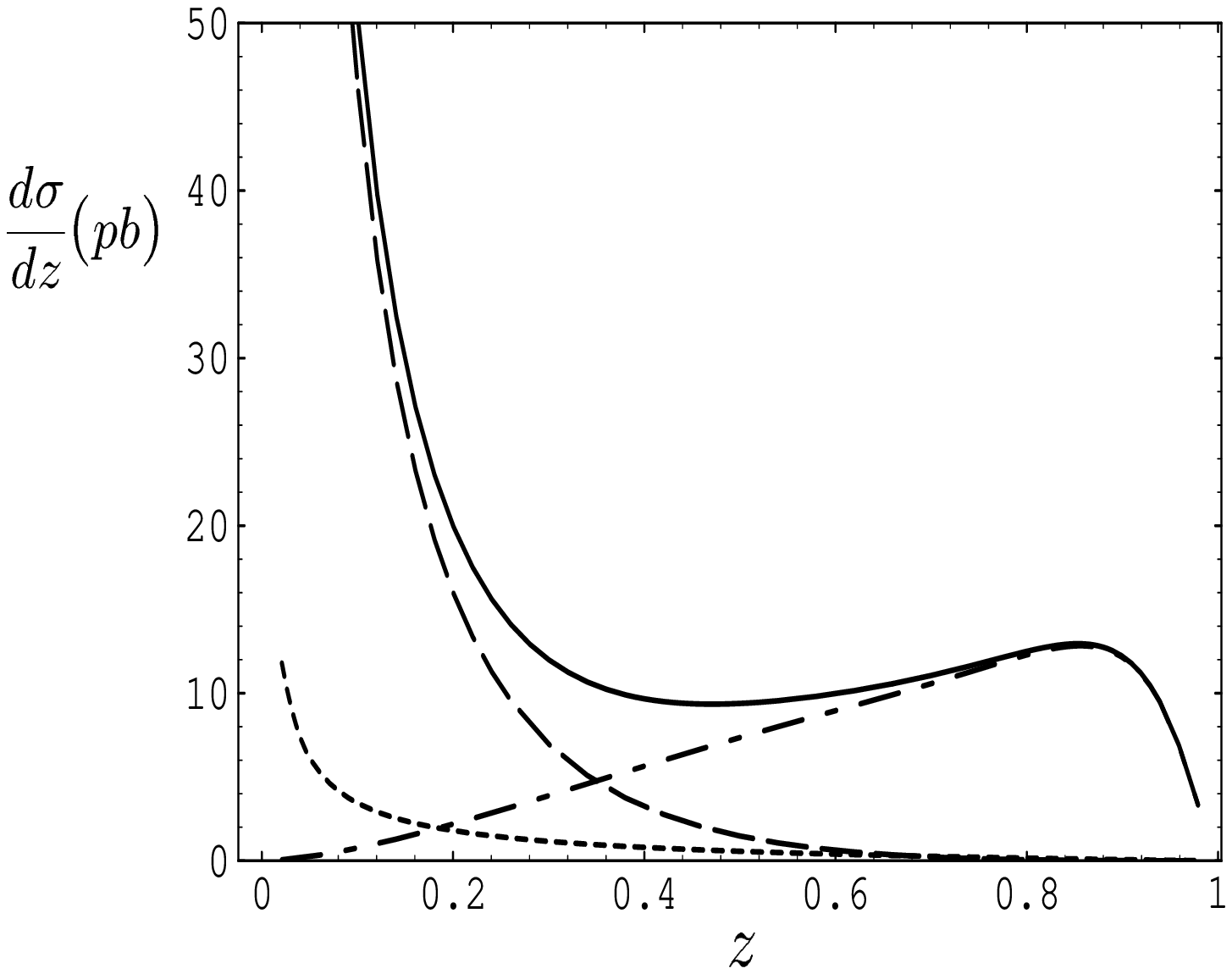}\hfill
\caption{} 
\label{NRQCDz}
\end{figure}

\begin{figure}
\epsfxsize=10cm
\hfil\epsfbox{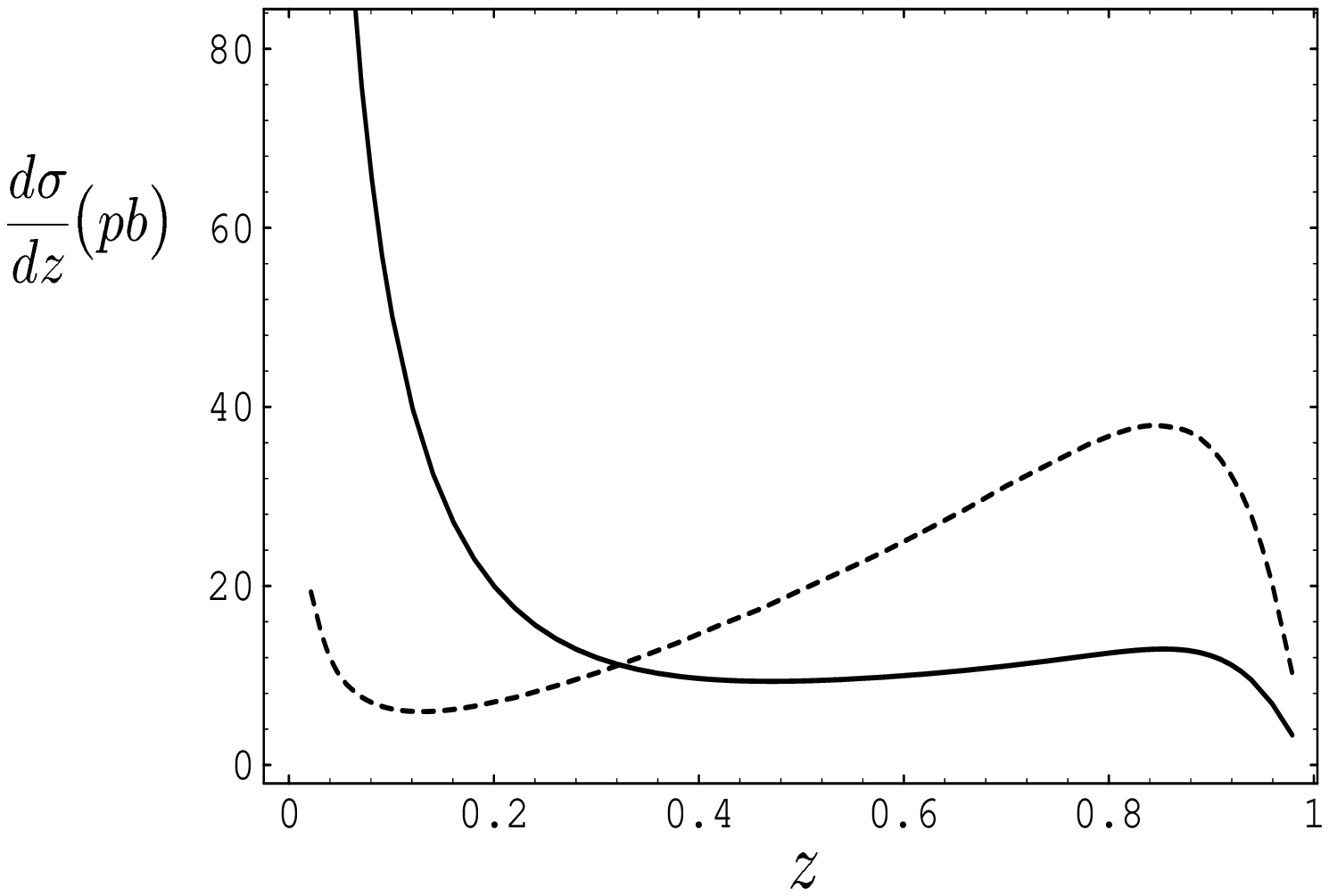}\hfill
\caption{} 
\label{z}
\end{figure}

More interesting are the distributions in the variable $z$. In fig.~\ref{NRQCDz}, I plot the NRQCD predictions for $d\sigma/dz$, at $\sqrt{s} = 100~{\rm GeV}$.  Color-singlet processes dominate at $z < 0.2$ and are negligible for $z > 0.5$, while color-octet processes give a significant enhancement of the cross section for $z > 0.4$. In fig.~\ref{z}, I compare the predictions of the CEM and the NRQCD factorization formalism. Since the CEM is dominated by a direct production process, the cross section in this model is
strongly peaked at high $z$. Resolved color-singlet production of $J/\psi$ dominates the NRQCD prediction for the cross section, leading to a substantial fraction of $J/\psi$ produced at low $z$, in stark contrast to the CEM where production at low $z$ is very small. While the color-octet component of the NRQCD prediction gives an enhancement in the large $z$ region, the NRQCD distribution is much flatter here than in the CEM.  

It is clear that the $z$ distribution in associated $J/\psi + \gamma$ production can be an excellent tool for distinguishing between different models of quarkonia production. An observation of associated $J/\psi + \gamma$ events with $z > 0.5$ would be a strong piece of evidence in favor of color-octet mechanisms. Observation of an enhancement of $J/\psi + \gamma$ events at low $z$ would rule out the CEM. Photoproduction of $J/\psi + \gamma$ also serves as an important new cross check of the extraction color-octet matrix elements at the Tevatron, since it is sensitive to the matrix element $\langle O^{J/\psi}_8(^3S_1) \rangle$, which is most important in fragmentation production of $J/\psi$ at large $p_{\perp}$. Photoproduction of $J/\psi + \gamma$ is complimentary to studies of low energy hadroproduction\cite{BR}, photoproduction\cite{AFM} and electroproduction \cite{S1}, which provide independent measurements of $\langle O^{J/\psi}_8(^1S_0) \rangle$ and $\langle O^{J/\psi}_8(^3P_0) \rangle$, but not $\langle O^{J/\psi}_8(^3S_1) \rangle$.

After this paper was submitted for publication, two preprints appeared which also discussed associated $J/\psi + \gamma$ production. Cacciari, Kr$\ddot{\rm{a}}$mer, and Greco \cite{CGK} present a detailed study of photoproduction as well as electroproduction of $J/\psi + \gamma$, including color-octet contributions. (A preliminary announcement of some of their results is given in a talk by Cacciari and Kr$\ddot{\rm{a}}$mer \cite{CK}.) Their results are in agreement with the results presented in this paper.
Early calculations of $J/\psi + \gamma$ production are reviewed in a recent preprint by Kim \cite{Kim}.  The calculations of $J/\psi + \gamma$ production in $ep$ collisions appearing in ref. \cite{Kim} are performed within the color-singlet model. Color-octet contributions may be important for some of the distributions presented in this paper. In particular, the photon rapidity distributions shown in fig.~9 of ref. \cite{Kim} were shown to recieve important color-octet corrections in ref. \cite{CGK}.

I would like to thank Bob Fletcher, Adam Falk and Sean Fleming for many enlightening discussions. This work was supported by the National Science Foundation under Grant No. PHY-9404057.

\begin{references}

\bibitem{CDF} F. Abe {\it et. al.}, Phys.\ Rev.\ Lett.\ {\bf 74} (1992) 3327.

\bibitem{BFY} E. Braaten, S. Fleming, and T. C. Yuan, OHSTPY-HEP-T-96-001, hep-ph/9602374.

\bibitem{Schuler} For an extensive review of the color-singlet model, see G. A. Schuler, CERN-TH-7170-94, hep-ph/9403387. 

\bibitem{Barbieri} R. Barbieri, R. Gatto and E. Remiddi, Phys.\ Lett.\ {\bf B61} (1976) 465; R. Barbieri, M. Caffo and E. Remiddi, Nucl.\ Phys.\ {\bf B162} (1980) 220; R. Barbieri, M. Caffo, R. Gatto and E. Remiddi, Phys.\ Lett.\ {\bf B95} (1980) 93; Nucl.\ Phys.\ {\bf B192} (1981) 61. 

\bibitem{BDFM} E. Braaten, M. A. Doncheski, S. Fleming and M. L. Mangano, Phys.\ Lett.\ {\bf B333} (1994) 548; D. P. Roy and K. Sridhar, Phys.\ Lett.\ {\bf B339} (1994) 141; M. Cacciari and M. Greco, Phys.\ Rev.\ Lett.\ {\bf 73} (1994) 1586;

\bibitem{BBL2} G.T. Bodwin, E. Braaten, and G.P. Lepage, Phys.\ Rev.\ {\bf D51} (1995) 1125 .

\bibitem{BF} E. Braaten and S. Fleming, Phys.\ Rev.\ Lett.\ {\bf 74} (1995) 3327.

\bibitem{ColorEvap} J.F Amundson, O.J.P. $\acute{\rm{E}}$boli, E. M. Gregores and F. Halzen, Phys.\ Lett.\ {\bf B372} (1996) 127; MADPH-96-942, hep-ph/9605295; O.J.P. $\acute{\rm{E}}$boli, E. M. Gregores and F. Halzen, MADPH-96-950, hep-ph/9607324; G. A. Schuler and R. Vogt, Phys.\ Lett.\ {\bf B387} (1996) 181. 

\bibitem{K} M. Kr$\ddot{\rm{a}}$mer, Nucl. Phys. {\bf B459} (1996) 3. 

\bibitem{BC} E. Braaten and Y. Chen, Phys.\ Rev. {\bf D54} (1996) 3216;
S. Fleming and I. Maksymyk, Phys.\ Rev. {\bf D54} (1996) 3608.

\bibitem{GK} J.F. Gunion and Z. Kunst, Phys.\ Lett.\ {\bf B178} (1986) 296.

\bibitem{EK} R.K. Ellis and Z. Kunst, Nucl.\ Phys.\ {\bf B303} (1988) 653.

\bibitem{KR} C.S. Kim and E. Reya, Phys.\ Lett.\ {\bf B300} (1993) 298.

\bibitem{GRVproton} M. Gl$\ddot{\rm{u}}$ck, E. Reya, and A. Vogt, Z.\ Phys.\ {\bf C67} (1995) 433.

\bibitem{GRVphoton} M. Gl$\ddot{\rm{u}}$ck, E. Reya, and A. Vogt, Phys.\ Rev.\ {\bf D46} (1992) 1973.

\bibitem{BSK} G.T. Bodwin, D. K. Sinclair and S. Kim, Phys.\ Rev.\ Lett.\ {\bf 77} (1996) 2376.

\bibitem{CL} P. Cho and A. Leibovich, Phys.\ Rev.\ {\bf D53} (1996) 150; Phys.\ Rev.\ {\bf D53} (1996) 6203.
 
\bibitem{S1} S. Fleming, MADPH-96-966, hep-ph/9610372; S. Fleming, I. Maksymyk, and T. Mehen, to be published.
 
\bibitem{BR} M. Beneke and I. Rothstein, Phys.\ Rev.\ {\bf D54}, (1996) 2005;
S. Gupta and R. Sridhar, Phys.\ Rev.\ {\bf D54} (1996) 5545; hep-ph/9608433;
W.K. Tang and M. V$\ddot{\rm{a}}$nttinen, Phys.\ Rev.\ {\bf D54} (1996) 4349;
Phys.\ Rev.\ {\bf D53} (1996) 4851.

\bibitem{AFM} J. Amundson, S. Fleming, and I. Maksymyk,  UTTG-10-95, hep-ph/9601298; M. Cacciari and M. Kr$\ddot{\rm{a}}$mer, Phys.\ Rev.\ Lett.\ {\bf 76} (1996) 4128; P. Ko, J. Lee and H.S. Song, Phys.\ Rev.\ {\bf D54} (1996) 4312.

\bibitem{CGK} M. Cacciari, M. Greco, and M. Kr$\ddot{\rm{a}}$mer, DESY 96-147, hep-ph/9611324.

\bibitem{CK} M. Cacciari and M. Kr$\ddot{\rm{a}}$mer, hep-ph/9609500.

\bibitem{Kim} C. S. Kim, KEK-TH-502, hep-ph/9612218.

\pagebreak

\noindent {\bf FIG. 1} - NRQCD predictions for $\gamma + p \rightarrow J/\psi + \gamma (+ X)$ total cross section as a function of $\sqrt{s}$. Dashed, dotted, and dashed-dotted lines indicate the contributions from color-singlet mechanisms, resolved color-octet mechanisms and  direct color-octet mechanisms, respectively. The sum of all production mechanisms is indicated by the solid line. 
\vspace{0.5 in}

\noindent {\bf FIG. 2} - Comparison of the NRQCD and CEM predictions for the total cross section. The solid line shows the NRQCD prediction, the dotted line shows the CEM prediction.
\vspace{0.5 in}

\noindent {\bf FIG. 3} - NRQCD predictions for $d \sigma/dz$ at $\sqrt{s} = 100~{\rm GeV}$. Dashed, dotted, and dashed-dotted lines indicate the contributions from color-singlet mechanisms, resolved color-octet mechanisms and  direct color-octet mechanisms, respectively. The sum of all production mechanisms is indicated by the solid line. 
\vspace{0.5 in}

\noindent {\bf FIG. 4} -  Comparison of NRQCD and CEM predictions for $d\sigma/dz$ at $\sqrt{s} = 100~{\rm GeV}$.  The solid line shows the NRQCD prediction, the dotted line shows the CEM prediction.
\vspace{0.5 in}
\pagebreak 


\pagebreak
\end{references}
\end{document}